\documentclass[12pt,a4paper,english]{article}
\usepackage{pslatex}
\usepackage[T1]{fontenc}
\usepackage[latin1]{inputenc}
\usepackage{longtable}
\usepackage{float}
\usepackage{amsmath}
\usepackage{graphicx}
\usepackage{amssymb}

\makeatletter

\providecommand{\tabularnewline}{\\}
\floatstyle{ruled}
\newfloat{algorithm}{tbp}{loa}
\floatname{algorithm}{Algorithm}

 \usepackage{algolyx}

\usepackage[top=2cm, left=2cm, right=2cm, bottom=3cm]{geometry}
\usepackage{cite}
\usepackage{amsthm}
\usepackage{algorithm}

\AtBeginDocument{
  
}

\usepackage{babel}
\makeatother
\begin{document}

\title{The Complexity of Simple Stochastic Games}

\author{Jonas Dieckelmann}

\maketitle
\begin{abstract}
In this paper we survey the computational time complexity of assorted
simple stochastic game problems, and we give an overview of the best
known algorithms associated with each problem.
\end{abstract}

\section{Introduction}

A simple stochastic game $\mathrm{G=(V,E)}$ is a directed graph whose
vertices are partitioned into four disjoint sets $V_{max}$, $V_{min}$,
$V_{avg}$ and $V_{sink}$. Depending on the set a vertex belongs
to, it is called max, min, average and sink vertex, respectively.
In addition, one of the vertices in $V$ is given the property of
being the start vertex. $V_{sink}$ contains exactly two vertices,
called the 1-sink and the 0-sink. The 1-sink and the 0-sink have no
children, while all other vertices have exactly two distinct children.
Loop edges $e=(i,i)$ are allowed. In the rest of this paper we assume
w.l.o.g. that $V=\{1,\ldots,n\}$ where $n-1$ is the 0-sink and $n$
is the 1-sink.

\begin{figure}[h]
\begin{centering}\includegraphics[bb=5bp 25bp 330bp 185bp,clip]{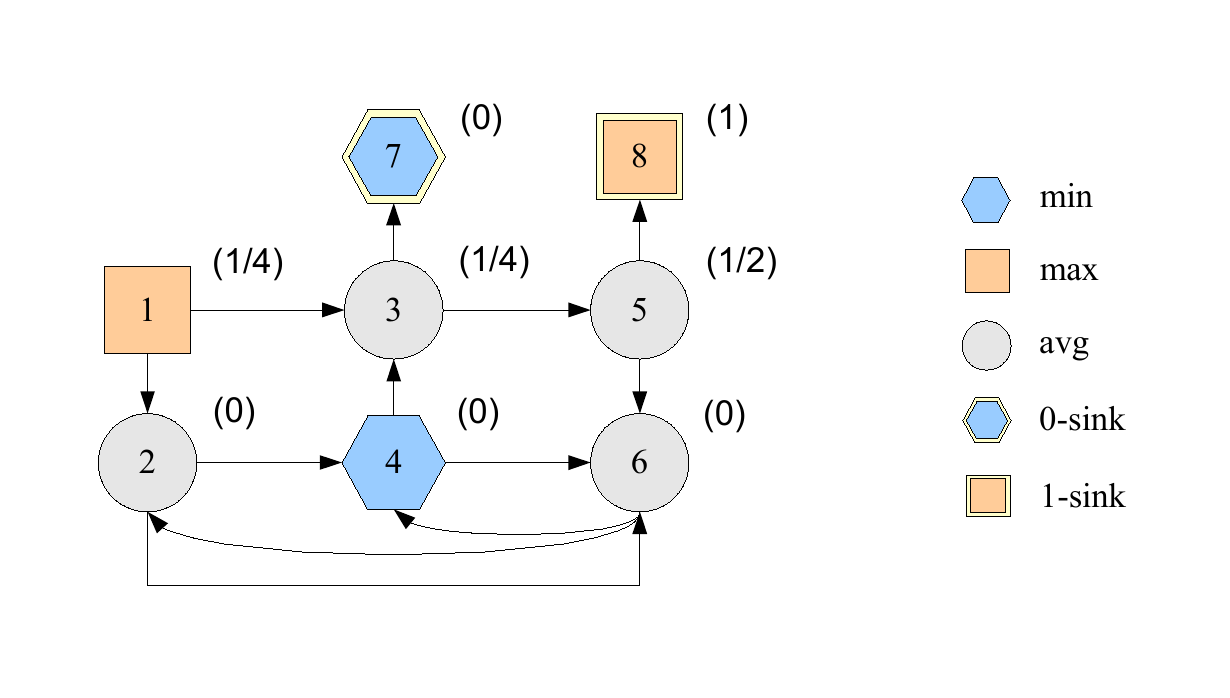}\label{fig:ssg}\par\end{centering}

\caption{A simple stochastic game with 8 vertices. Vertex 1 is the start vertex.
The numbers in parentheses denote the optimal vertex values. }
\end{figure}

The game is played by two players, called the max player and the min
player, who have diametrically opposed objectives. At the start of
the game, a token is placed on the start vertex. In each round, the
token is moved from a vertex to one of its children obeying the following
rule: whenever the token is positioned on a max vertex, the max player
decides to which child the token is moved; whenever the token is positioned
on a min vertex, the min player decides to which child the token is
moved and whenever the token is positioned on an average vertex, the
token is moved with probability $1/2$ to one of its children. Average
vertices hence model randomness in this kind of stochastic game. The
game ends, when the token reaches a sink vertex. The max player wins
the game, if the token reaches the 1-sink. The min player wins the
game in all other cases -- that is either when the token reaches the
0-sink or when he can force an infinite play where the token reaches
neither sink vertex.

The optimal value of a vertex is defined as the probability that the
max player wins the game starting at that vertex, assuming both players
employ optimal strategies (a strategy is optimal, if the probability
of winning the game with it is greater or equal to that of any other
strategy, regardless of the strategy chosen by the opponent). We will
later see that both players of a simple stochastic game posses optimal
strategies, albeit not unique ones. The value of the game is defined
to be the optimal value of its start vertex, and the optimal value
vector of the game is defined to be the vector whose components are
the optimal vertex values of the game.

The most intriguing question to be asked about a simple stochastic
game is: what is its value? As we shall see later, the complexity%
\footnote{We will refer to the computational time complexity of a problem as
the problem's complexity.%
} of the associated function problem is polynomial-time equivalent
to that of finding the optimal value vector of the game. The problem
of computing the optimal value vector of a simple stochastic game
has been studied extensively from an algorithmic point of view \cite{Condon93,Somla05,BjorklundV05,Kumar04,Ludwig95,SzepesvL96}
(no polynomial time algorithm has been found), but the author is not
aware of any previous efforts in studying its complexity. Therefore,
we shall prove containment of the problem in FNP in a subsequent section
of this paper. 

The question about a simple stochastic game's value is not only intriguing,
it has also pratical relevance. This is because stochastic games are
nowadays used as a formal tool in a variety of different application
areas, including automated software verification and controler optimization,
where the game's value constitutes the single most crucial information.
Apart from this, there exist some other motivations behind the study
of simple stochastic games. Most of these are related to the \emph{SSG-VALUE}
problem -- given a simple stochastic game, is its value greater than
$1/2$? Though Condon \cite{Condon92} was able to show that the \emph{SSG-VALUE}
problem is contained in $NP\cap coNP$, despite significant efforts
\cite{Juba06,GartnerR05,BjorklundV05} to obtain a hardness result
for a specific complexity class, the problem's exact complexity status
is unknown. Thus one of the motivations behind the study of simple
stochastic games is the desire to find a complexity class for which
\emph{SSG-VALUE} is complete, so as to obtain a clue whether the problem
is intractable%
\footnote{Intractable computational problems are those that feature exponential
time or space complexities.%
} or not. The present consensus is that, since contained in $NP\cap coNP$\emph{,
SSG-VALUE} is very likely not NP-complete and may allow for more efficient
algorithms than the exponential ones currently known. Condon reinforces
this hypothesis by stating that \emph{SSG-VALUE} constitutes one of
the rare \emph{combinatorial} problems to be contained in $NP\cap coNP$,
but for which containment also in P is an open question. 

A last motivation behind the study of simple stochastic games can
be expressed as the {}``kill two birds with one stone'' factor;
many computational problems, such as the generalized linear complementarity
problem \emph{(GLCP)} and the minimum stable circuit problem for min/max/avg-circuits
\emph{(STABLE-CIRCUIT)}, were shown \emph{\cite{GartnerR05,Juba06}}
to be polynomial-time reducible to \emph{SSG-VALUE --} hence more
efficient algorithms for \emph{SSG-VALUE} will also yield more efficient
algorithms for those other problems. The rest of this paper is organized
as follows: in section 2, we will restate the essential definitions
for simple stochastic games as given in Condon's initial paper on
the subject. In section 3, we will provide a more detailed view of
simple stochastic games which will enable us to conduct our complexity
survey in section 4. The paper concludes with a summary of the important
points and an overview of open problems in section 5.

\section{Definitions}

\subsection{Player Strategies}

Given a simple stochastic game, a strategy $\tau$ for the min player
(or min strategy) is a subset of the game's edges such that for each
min vertex $i$ with children $j$ and $k$, either $(i,j)\in\tau$
or $(i,k)\in\tau$ applies. Substituting $\tau$ with $\sigma$ and
min with max in the above sentence, we obtain the analog definition
for the max strategy $\sigma$. Informally, a strategy denotes the
player's choice to which child the token is to be moved whenever it
is positioned on a vertex belonging to that player. The reason for
defining player strategies like this will be explained in the next
section of this paper.

\subsection{Reduced Games}

Given a simple stochastic game $G=(V,E)$ and a strategy $\tau$ to
be employed by the min player, the reduced game $G_{\tau}$ is defined
to be the sub-graph of $G$ obtained by removing all edges from $G$
that are not selected by the min strategy $\tau$, i.\,e.\[
G_{\tau}=(V,E_{\tau})\quad\text{where}\quad E_{\tau}=E\setminus\{(i,j)\in E:i\in V_{min}\wedge(i,j)\notin\tau\}\]
$G_{\tau}$ can be regarded as the 1-player equivalent of $G$, where
it is certain that the min player employs $\tau$. In a similar manner,
the reduced games $G_{\sigma}$ and $G_{\tau,\sigma}$ are defined
as\[
G_{\sigma}=(V,E_{\sigma})\quad\text{where}\quad E_{\sigma}=E\setminus\{(i,j)\in E:i\in V_{max}\wedge(i,j)\notin\sigma\}\]
\[
G_{\tau,\sigma}=(V,E_{\tau,\sigma})\quad\text{where}\quad E_{\tau,\sigma}=E\setminus\{(i,j)\in E:i\in V_{min}\cup V_{max}\wedge(i,j)\notin\tau\cup\sigma\}\]

We observe that in the reduced game $G_{\tau,\sigma}$, the strategies
of both players are fixed to $\tau$ and $\sigma$ and the winner
is decided by a (more or less) random walk of the token on the graph
of $G_{\tau,\sigma}$.

\subsection{Vertex Values}

Given a reduced game $G_{\tau,\sigma}$ (or alternatively a simple
stochastic game $G$ and a pair of strategies $\tau$ and $\sigma$
to be employed by the players), the value of vertex $i$, $v_{\tau,\sigma}(i)$,
is defined to be the probability that the token reaches the 1-sink
in a random walk on the graph of $G_{\tau,\sigma}$, starting at vertex
$i$. The value vector $\vec{v}_{\tau,\sigma}$ of $G_{\tau,\sigma}$
is defined to be the vector whose components are the vertex values
of $G_{\tau,\sigma}$.

\subsection{Optimal Player Strategies and Optimal Vertex Values}

Given a simple stochastic game $G$, a strategy $\tau_{opt}$ for
the min player satisfying\[
v_{\tau_{opt},\sigma_{opt}}(i)\leq v_{\tau,\sigma_{opt}}(i)\quad\forall i,\tau\]
is called an optimal strategy for the min player. Similarly, a strategy
$\sigma_{opt}$ for the max player satisfying\[
v_{\tau_{opt},\sigma_{opt}}(i)\geq v_{\tau_{opt},\sigma}(i)\quad\forall i,\sigma\]

is called an optimal strategy for the max player. Informally, the
formulas say that by employing an optimal strategy for $G$, a player
assures himself the highest probability of winning $G$ no matter
what the start vertex. 

The \emph{optimal} value of vertex i, $v(i)$, is defined as\[
v(i)=v_{\tau_{opt},\sigma_{opt}}(i)\]
where $\tau_{opt}$ and $\sigma_{opt}$ are a pair of optimal player
strategies for $G$. The optimal value of a vertex denotes the probability
that the max player wins the game starting at that vertex, assuming
both players employ optimal strategies. The optimal value vector $\vec{v}$
of $G$ is defined to be the vector whose components are the optimal
vertex values of $G$. Misleadingly, the value of a simple stochastic
game is the \emph{optimal} (and not just any) value of the start vertex.
It is also important not to confuse vertex values with optimal vertex
values, as their meaning is different.

\subsection{Stopping Simple Stochastic Games}

A stopping simple stochastic game is a simple stochastic game which
does not permit infinite plays, i.\,e. the token always reaches a
sink vertex after a finite number of rounds, regardless of the strategies
chosen by the players. More precisely, if for all pairs of strategies
$\tau,\sigma$ each vertex of the reduced game $G_{\tau,\sigma}$
has a path to a sink vertex, then $G$ is stopping. Stopping stochastic
games are also referred to as stochastic games that halt with probability
1.

\section{Properties of Simple Stochastic Games}

From the introduction, we can already derive some important \emph{game
theoretic} properties of simple stochastic games. We will use these
in the elaborations to follow:

\begin{itemize}
\item determined -- the optimal value vector exists and is unique.
\item finite -- the game has n states, and in each state the players have
at most two actions to choose from.
\item zero sum -- in every state of the game, the win expectancy for one
player is the complement of the win expectancy for the opponent.
\item perfect information -- the players act sequentially and each player
is completely informed about the history and the state of the game.
\item reachability objective -- the objective of the players is to force
the token to reach their respective sink vertex.
\item $2\frac{1}{2}$ player -- the coin-flipping ruler over average vertices
(nature) is given the status of a half player.
\end{itemize}
Let us start by discussing player strategies for simple stochastic
games. The rationale of defining player strategies just as we did
originates from the initial work on stochastic games by Shapley \cite{Shapley53};
he showed that perfect information stopping stochastic games -- and
hence stopping simple stochastic games -- have a Nash equilibrium
\cite{ChatterjeeMJ04} in pure%
\footnote{a strategy is called pure if it consists of deterministic (as opposed
to random) choices by the player.%
} memoryless%
\footnote{a strategy is called memoryless if the players choice only depends
on the state of the game, and not its history.%
} optimal strategies (Somla \cite{Somla05} points out in a footnote
that it is possible to extend this results to non-stopping simple
stochastic games, using advanced proof techniques). Because of this
fact, it suffices to denote a player's strategy by a prescription
that, for each vertex belonging to the player, states to which child
the token is to be moved; such a prescription can be modeled as a
subset of the game's edges.

Furthermore, as a direct consequence of the definition of optimal
strategies, we find that a min strategy is optimal for a particular
game $G$ if and only if it is locally optimal (or greedy) at every
min vertex of $G$ with respect to the optimal value vector. That
is to say, the best strategy for the min player is to always move
the token from a min vertex to the child which has got the lower optimal
value. The same statement can be made for the max player, but certainly,
the max player always moves the token from a max vertex to the child
which posesses the higher optimal value. We conclude that a player
cannot improve his performance by making local concessions -- unlike
in chess, non-greediness will not be rewarded.

Another property of optimal player strategies is that they are not
necessarily unique; instead, a simple stochastic game may posses more
than one optimal strategy for a player. As a trivial example, picture
a simple stochastic game which contains a min vertex that has itself
and the 0-sink as children. In this game, the optimal value of the
min vertex is 0 and the min player possesses at least two different
optimal strategies for the game -- one of which contains the edge
to the 0-sink and one of which contains the loop edge.

We have mentioned that player strategies are independent of the game's
history and deduce that they can be fixed before the start of the
game. Once both players have fixed their strategies to be $\tau$
and $\sigma$, a random walk of the token on the reduced game $G_{\tau,\sigma}$
decides upon the winner. For the upcoming discussion about the reduced
game $G_{\tau,\sigma}$ let us w.l.o.g. assume that the vertices in
$G_{\tau,\sigma}$ are labeled in such a way that the vertices $1,\ldots,t$
are those that have a path to a sink vertex. With this in mind, we
can easily verify that -- following its definition -- the value vector
$\vec{v}_{\tau,\sigma}$ of $G_{\tau,\sigma}$ is a solution to the
following system of linear equations: $v_{\tau,\sigma}(n)=1$, $v_{\tau,\sigma}(i)=0$
for $t<i<n$ and otherwise\[
v_{\tau,\sigma}(i)=\begin{cases}
v_{\tau,\sigma}(j) & \text{if }i\text{ is a min or max vertex with child }j\\
\frac{1}{2}(v_{\tau,\sigma}(j)+v_{\tau,\sigma}(k)) & \text{if }i\text{ is an average vertex with children }j\text{ and }k\end{cases}\]

which can be written as\begin{equation}
\vec{v}_{\tau,\sigma}=Q\vec{v}_{\tau,\sigma}+\vec{b}\qquad\Leftrightarrow\qquad(I-Q)\vec{v}_{\tau,\sigma}=\vec{b}\label{eq:valuevector}\end{equation}

where $Q\in\mathbb{Q}^{n\times n}$ is related to the topology of
$G_{\tau,\sigma}$ as follows: $Q_{ij}=0$ if $i>t$ and otherwise\begin{flalign*}
Q_{ij} & =\begin{cases}
1 & \text{if }i\text{ is a min or max vertex with child }j\\
\frac{1}{2} & \text{if }i\text{ is an average vertex with child }j\end{cases}\end{flalign*}

and $\vec{b}\in\mathbb{Q}^{n}$ is defined as\[
b_{i}=\begin{cases}
1 & \text{if }i=n\\
0 & \text{otherwise}\end{cases}\]

We will rewrite the proof given by Condon \cite{Condon92} that \eqref{eq:valuevector}
has a unique solution. Let $\lambda_{i}$ be the $i$-th eigenvalue
of $Q$. The idea is to show that as $m\rightarrow\infty$, $Q^{mn}\rightarrow0$,
from which the following chain of deductions can be made\begin{eqnarray*}
Q^{mn}\rightarrow0 & \Rightarrow & \lambda_{i}\neq1\qquad i=1\ldots n\\
 & \Leftrightarrow & det(Q-I)\neq0\\
 & \Leftrightarrow & rank(Q-I)=n\\
 & \Leftrightarrow & (1)\text{ has a unique solution}\end{eqnarray*}

Let us denote the upper t rows of $Q$ by the $t\times n$ matrix
$Q_{t}$, which is the 1-step transition matrix of non-sink vertices
of $G_{\tau,\sigma}$ that have a path to a sink vertex. As a matter
of fact, entry $ij$ of $Q_{t}^{mn}=Q_{t}\cdots Q_{t}$ denotes the
probability that the token reaches vertex $j$ from vertex $i$ in
a random walk on the graph $G_{\tau,\sigma}$ in exactly $mn$ steps.
Therefore, the sum of values in the $i$-th row of $Q_{t}^{mn}$ equals
one minus the probability of reaching a sink vertex from $i$ in $k<mn$
steps. The probability of reaching a sink vertex from $i$ in $k<n$
steps is greater than zero, as $i$ has at least one path to a sink
vertex of length no more than the maximal diameter of $G_{\tau,\sigma}$
-- which is $n-1$. Additionally, for $m'>m$, the probability of
reaching a sink vertex from $i$ in $k<m'n$ steps is obviously greater
than the probability of reaching a sink vertex from $i$ in $k<mn$
steps. As the values of $Q_{t}^{mn}$ are all positive, it follows
that as $m\rightarrow\infty$, $Q_{t}^{mn}\rightarrow0$ and thus
$Q^{mn}\rightarrow0$. 

Using a local graph search algorithm, one is able to verify in time
$O(n)$ whether a given vertex of $G_{\tau,\sigma}$ has a path to
a sink vertex. Therefore, $Q$ and $\vec{b}$ can be constructed from
$G_{\tau,\sigma}$ in time $O(n^{2})$. By solving \eqref{eq:valuevector}
with Gauss-elimination or $LU$-decomposition (both $O(n^{3})$),
we obtain a cubic time algorithm for the problem of computing the
value vector of $G_{\tau,\sigma}$. 

In a related concern, Condon \cite{Condon92} showed that the vertex
values $v_{\tau,\sigma}(i)$ of the reduced game $G_{\tau,\sigma}$
are rational numbers from the set\begin{equation}
\Omega_{t}=\{ p/q\in\mathbb{Q}:0\leq p\leq q\leq4^{t}\}\label{eq:omega}\end{equation}
where $t$ is defined as before, i.\,e. $t$ is the number of non-sink
vertices of $G_{\tau,\sigma}$ which have a path to a sink vertex.
To understand why this is true, consider that, as $\vec{v}_{\tau,\sigma}$
is a solution to \eqref{eq:valuevector}, the components of $\vec{v}_{\tau,\sigma}$
can be denoted by $v_{\tau,\sigma}(i)=D_{i}/D$ where $D$ is the
determinant of the matrix $I-Q$ and $D_{i}$ is the determinant of
$I-Q$ which has the i-th column replaced by $\vec{b}$. Since the
components of $I-Q$ and $\vec{b}$ are all rational, both $D_{i}$
and $D$ must also be rational. Condon concludes the proof by showing
that $0\leq D_{i}\leq D\leq4^{t}$ holds. Note that not all values
from the set $\Omega_{t}$ can occur as vertex values in the game
$G_{\tau,\sigma}$; instead, $\Omega_{t}$ is a superset -- or approximation
-- of the possible vertex values of $G_{\tau,\sigma}$. In the context
of finding a reduced game's optimal vertex values, $\Omega_{t}$ can
be regarded as a search space of that problem.

\begin{figure}[H]
\begin{centering}\includegraphics[bb=20bp 10bp 490bp 95bp,clip]{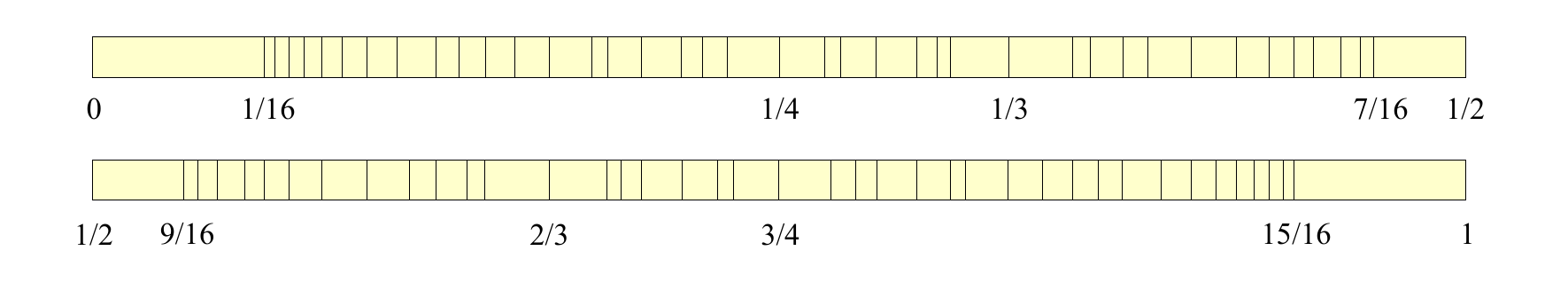}\par\end{centering}

\caption{Visualization of the set $\Omega_{2}$. The possible vertex values
of a simple stochastic game with 4 vertices are a subset of the depicted
values.}
\end{figure}

Concluding the reduced game topic, it is worth mentioning that reduced
games can also be studied in the framework of Markov processes. If
we were to assign transition probabilities of 1 to the single edges
leaving player vertices in $G_{\sigma,\tau}$, next to the already
established transition probabilities of $1/2$ assigned to the edges
leaving average vertices, then the so modified graph $G_{\sigma,\tau}$
would formally conform to the definition of a Markov chain. In a similar
fashion, the reduced games $G_{\tau}$ and $G_{\sigma}$ can be transformed
into Markov decision processes (MDP's). 

Following its definition, the optimal value vector $\vec{v}$ of a
simple stochastic game $G$ is a solution to the equation system\[
v(i)=\begin{cases}
\max(v(j),v(k)) & \text{if }i\text{ is a max vertex with children }j\text{ and }k\\
\min(v(j),v(k)) & \text{if }i\text{ is a min vertex with children }j\text{ and }k\\
\frac{1}{2}(v(j)+v(k)) & \text{if }i\text{ is an average vertex with children }j\text{ and }k\\
0 & \text{if }i=n-1\\
1 & \text{if }i=n\end{cases}\]

which can be written as\begin{equation}
\vec{v}=I_{G}(\vec{v}\mathrm{)}\label{eq:optimalvaluevector}\end{equation}

for $I_{G}:[0,1]^{n}\rightarrow[0,1]^{n}$ as defined above. Contrary
to \eqref{eq:valuevector}, the equations in \eqref{eq:optimalvaluevector}
are non-linear and a solution can no longer be derived analytically
but rather has to be computed numerically. Shapley \cite{Shapley53}
showed that in the case of stopping stochastic games -- and hence
stopping simple stochastic games -- the operator $I_{G}$ is contracting
on the hypercube $[0,1]^{n}$ and therefore has a unique fixed point.
In this case, the solution to \eqref{eq:optimalvaluevector} is the
optimal value vector of the game. In the case of non-stopping simple
stochastic games however, \eqref{eq:optimalvaluevector} is necessary,
but not sufficient, for the optimal value vector of the game. For
an example of ambiguous values in a simple stochastic game that has
only one connected component and is non-trivial observe figure \ref{fig:ssg}.
In the depicted game, any value below $1/4$ can be assigned uniformly
to the vertices $\{2,4,6\}$ without violating \eqref{eq:optimalvaluevector}
-- though only 0 is the correct optimal value for each of the vertices.
We finish this paragraph with the statement that, contrary to optimal
strategies, the optimal value vector \emph{is} unique in every simple
stochastic game. 

\begin{figure}[h]
\begin{centering}\includegraphics[bb=10bp 16bp 315bp 92bp,clip]{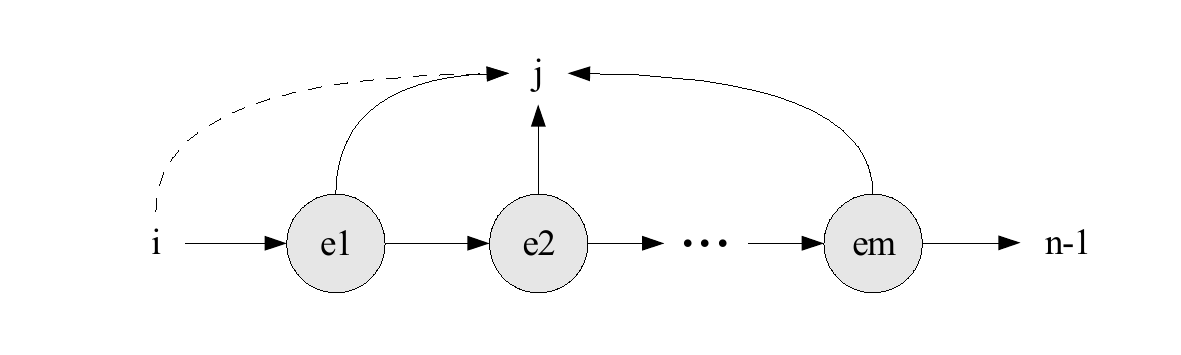}\label{fig:betassg}\par\end{centering}

\caption{Construction rule for the $\beta$-stopping game $G'$ of $G$, where
$\beta=1/2^{m}$ and $m=cn$. The dashed edge is present in $G$,
but not in $G'$; it is replaced by the depicted elements.}
\end{figure}

The last discovery about simple stochastic games which is relevant
in the context of this paper is again due to Condon \cite{Condon92}.
She found out that from a simple stochastic game $G$, a stopping
simple stochastic game $G'$ can be constructed whose vertex values
are arbitrarily close to the vertex values of $G$. The construction
rule is as follows: For $\beta=1/2^{cn}$, the so-called $\beta$-stopping
game $G'$ adopts all the vertices of $G$ but it does not adopt any
edge of $G$. For each edge $(i,j)$ of $G$, the graph $G'$ instead
contains a path of $m=cn$ average vertices which are connected to
the vertices $i$, $j$ and $n-1$ (the 0-sink) as depicted in figure
\ref{fig:betassg}.

Two important properties of the $\beta$-stopping game $G'$ have
been set forth by Condon:

\begin{enumerate}
\item $G'$ can be constructed from $G$ in time $O(n^{2})$, where the
constant $c$ only has a linear effect on the runtime of the construction
algorithm and is hidden in the O-notation. 
\item For arbitrary player strategies $\tau$ and $\sigma$, the corresponding
value vectors $\vec{v}_{\tau,\sigma}$ of $G$ and ${\vec{v}'}_{\tau,\sigma}$
of $G'$ satisfy \begin{equation}
|v_{\tau,\sigma}(i)-v'_{\tau,\sigma}(i)|\leq2^{n(3-c)}\qquad i\in V\label{eq:closevalues}\end{equation}

\end{enumerate}
We observe that by choosing $c$ large enough, the differences between
corresponding vertex values of the games $G$ and $G'$ can be made
arbitrarily small, though $G'$ can still be constructed in time polynomial
in the size of $G$. We will use this result, as well as previous
results from this section, for proving the claims we make in the next
section.

\section{Complexity Survey}

We begin this section by showing that -- given a simple stochastic
game $G$ -- the below function problems have polynomial-time equivalent
complexities:

\begin{enumerate}
\item What is the value of $G$?
\item What is the optimal value vector of $G$?
\item What are optimal player strategies of $G$? 
\end{enumerate}
\begin{proof}
 {}``$2\leq_{p}1$'': Let us assume that Algorithm $A1$ computes
the value of the game $G$. From $A1$, we construct an Algorithm
$A2$ which computes the optimal value vector of $G$. On input of
$G$, $A2$ iterates over all vertices $i$ of $G$. In each iteration,
$A2$ changes the start vertex of $G$ to be $i$ and then performs
a run of $A1$ on $G$, yielding the optimal vertex value $v(i)$.
After the last iteration, $A2$ outputs the computed values in form
of the optimal value vector of $G$. The runtime of $A2$ is dominated
by the queries to $A1$ and since $A2$ makes n queries to $A1$,
$A2$ is efficient given $A1$ is.

{}``$3\leq_{p}2$'': Suppose Algorithm $A2$ computes the optimal
value vector of the game $G$. We give an informal description of
an Algorithm $A3$ which, using $A2$, computes the optimal player
strategies of $G$. Our tool will be the result from the previous
section that every min (max) strategy, which is locally optimal at
every min (max) vertex of $G$, is an optimal min (max) strategy of
$G$; it therefore suffices for $A3$ to compute two locally optimal
strategies $\tau_{opt}$ and $\sigma_{opt}$ for the players. On input
of $G$, $A3$ runs $A2$ on $G$, yielding the optimal value vector
$\vec{v}$ of $G$. For each min vertex $i$ with children $j$ and
$k$, $A3$ adds $(i,j)$ to the initial empty $\tau$ if $v(j)\leq v(k)$,
else $A3$ adds $(i,k)$ to $\tau$. Similarly, for each max vertex
$i$ with children $j$ and $k$, $A3$ adds $(i,j)$ to the initial
empty $\sigma$ if $v(j)\geq v(k)$, else $A3$ adds $(i,k)$ to $\sigma$.
Following this construction, $\tau_{opt}$ and $\sigma_{opt}$ are
locally optimal strategies and since $A3$ makes one query to $A2$
and performs $O(n)$ instructions, $A3$ is efficient given $A2$
is.

{}``$1\leq_{p}3$'': From an algorithm $A3$ which computes a pair
of optimal player strategies of the game $G$, we construct an algorithm
$A1$ which computes the value of $G$. On input of $G$, $A1$ runs
$A3$ on $G$ to obtain the optimal player strategies $\tau_{opt}$
and $\sigma_{opt}$ of $G$. In time $O(n)$, $A1$ then constructs
the reduced game $G_{\tau_{opt},\sigma_{opt}}$ corresponding to $\tau_{opt}$
and $\sigma_{opt}$. We already argued in the previous section that
the value vector (and hence the value) of a reduced game can be computed
in time polynomial in the size of the game. Since $A1$ makes one
call to $A3$, $A1$ is efficient given $A3$ is.\end{proof}

\subsubsection*{\emph{SSG-TWOKIND} \textmd{(function)}}

\begin{longtable}{|ll|}
\hline 
Input:&
Simple stochastic game $G$, lacking one vertex kind\tabularnewline
Question:&
What is the optimal value vector $\vec{v}$ of $G$?\tabularnewline
\hline
\end{longtable}

\textbf{\emph{Complexity}} \textbf{}\textbf{\emph{class:}} FP

\textbf{\emph{Algorithms:}} 

Papers discussing algorithms for \emph{SSG-TWOKIND} include \cite{Derman70,Condon93,FilarSTV91,SzepesvL96,Somla05,Andersson06}.
In the case that $G$ lacks min vertices, \emph{SSG-TWOKIND} can be
expressed as the following linear optimization (or programming) problem,
as was first shown by Derman \cite{Derman70}\[
\sum_{i=1}^{n}v(i)\rightarrow\min\]
subject to $v(n-1)=0$, $v(n)=1$ and\[
v(i)\geq\begin{cases}
0 & 1\leq i\leq n\\
v(j) & \text{if }i\text{ is a max vertex with child }j\\
\frac{1}{2}(v(j)+v(k)) & \text{if }i\text{ is an average vertex with children }j\text{ and }k\end{cases}\]
Similarly, in the case that $G$ lacks max vertices, \emph{}Derman
showed that the optimal value vector of $G$ is the unique solution
to the linear optimization problem \[
\sum_{i=1}^{n}v(i)\rightarrow\max\]
subject to $v(n-1)=0$, $v(n)=1$ and\[
v(i)\leq\begin{cases}
0 & 1\leq i\leq n\\
v(j) & \text{if }i\text{ is a min vertex with child }j\\
\frac{1}{2}(v(j)+v(k)) & \text{if }i\text{ is an average vertex with children }j\text{ and }k\end{cases}\]

Khachian \cite{Khachian79} was the first to show that linear programming
problems can be solved in time polynomial in the bits needed to describe
the problem. Since the amount of bits needed to encode any of the
above formulas is a polynomial function of $n$ \cite{Condon92},
we obtain the fruit that the value vector $\vec{v}$ of $G$ can be
computed in polynomial time. 

So called interior point algorithms for linear programming problems,
which can move \emph{through} the feasible region%
\footnote{The feasible region of a linear programming problem is the set of
variable evaluations (vertices) which satisfy the constraints given
in the problem.%
} instead than just along its boundary, perform best in practice. According
to \href{http://en.wikipedia.org}{wikipedia}, Mehrotra's \cite{Mehrotra91}
interior point algorithm is regarded as the fastest, though a worst
case boundary is not available. Up to date, there exists no strongly
polynomial time algorithm for linear programming problems, i.e one
that is polynomial in the number of variables of the problem up to
order \textasciitilde{}4. 

In the case that $G$ lacks average vertices, the algorithm given
in appendix A correctly computes the optimal value vector of $G$.
The number of executions of the repeat loop is $O(n)$, as $D$ is
static after at most $n-2$ executions. It follows that the algorithm
has quadratic runtime, which is a much better result as that obtained
for the above linear programming problems.

\subsubsection*{\emph{SSG-OVV} \textmd{(function)}}

\begin{longtable}{|ll|}
\hline 
Input:&
Simple stochastic game $G$\tabularnewline
Question:&
What is the optimal value vector $\vec{v}$ of $G$?\tabularnewline
\hline
\end{longtable}

\textbf{\emph{Complexity}} \textbf{}\textbf{\emph{class:}} FNP

We will first give an informal description of the proof that \emph{SSG-OVV}$\in FNP$.
Let $\Omega_{n}\supset\Omega_{t}$ be a superset of the possible vertex
values of $G$, as discussed in \eqref{eq:omega}. The proof is based
on the fact that there exists exactly one vector $\vec{z}$ from the
set $\Omega_{n}^{n}$ -- namely the optimal value vector of $G$ --
which satisfies $|z(i)-v'(i)|\leq4^{-2n}/2$ for all $i\in V$, where
$\vec{v}'$ is the optimal value vector of the $1/2^{9n}$-stopping
game $G'$ of $G$. We deduce that a nondeterministic Turing-machine
$M$ for problem \emph{SSG-OVV} could guess an arbitrary vector $\vec{z}\in\Omega_{n}^{n}$
and -- assuming $M$ knows $\vec{v}'$ -- argue in polynomial time
that $\vec{z}=\vec{v}$ if and only if $\vec{z}$ satisfies the mentioned
constraint. Of course, $M$ cannot compute $\vec{v}'$ from scratch,
but since $M$ is nondeterministic, we can happily let it, next to
$\vec{v}$, also guess $\vec{v}'$. By evaluating \eqref{eq:optimalvaluevector},
$M$ can easily verify whether its guess of $\vec{v}'$ is correct.
Therefore, letting $M$ guess the optimal value vectors of both $G$
and $G'$, $M$ is able to conduct a polynomial time verification
of both guesses.

Considering a formal proof, we first show that the difference between
two vertex values of $G$ is either 0 or has a lower bound $\delta$.
Let $\alpha,\beta\in\Omega_{n}$. Then either $|\alpha-\beta|=0$
or \[
|\alpha-\beta|=\left|\frac{p}{q}-\frac{p'}{q'}\right|=\left|\frac{pq'-p'q}{qq'}\right|\geq\frac{1}{qq'}\geq\frac{1}{4^{n}4^{n}}=4^{-2n}=:\delta\]

Following \eqref{eq:closevalues}, we further observe that the differences
between corresponding vertex values in the games $G$ and the associated
$1/2^{9n}$-stopping game $G'$ are bound below $\delta/2$, i.\,e.
for arbitrary strategies $\tau$ and $\sigma$ \[
|v_{\tau,\sigma}(i)-{v'}_{\tau,\sigma}(i)|\leq2^{-6n}=4^{-3n}<\delta/2\qquad i\in V\]

On input of $G$, a nondeterministic Turing-machine $M$ for problem
\emph{SSG-OVV} first guesses one vector from the set $\Omega_{n}^{n}$
to be the optimal value vector of $G$ and one vector from the set
$\Omega_{n'}^{n'}$ to be the optimal value vector of $G'$, where
$n'=9n|E|+n$. Let us denote these vectors by the tuple $(\vec{z},\vec{s})$.
If we define that $M$ accepts $(\vec{z},\vec{s})$ if $\vec{z}=I_{G}(\vec{z})$,
$\vec{s}=I_{G'}(\vec{s})$ and $|z(i)-s(i)|<\delta/2$ for all $i\in V$,
then $M$ accepts $(\vec{z},\vec{s})$ if and only if $\vec{z}=\vec{v}\wedge\vec{s}=\vec{v}'$.
Also it should be clear that $M$ comes to a conclusion in time polynomial
in the number of vertices of $G$, as in particular $M$ is able to
construct $G'$ from $G$ in time $O(n^{2})$. 

\begin{proof}
 Instead of saying {}``$M$ accepts $(\vec{z},\vec{s})$'', we will
just say {}``$M$ accepts''. {}``$\Rightarrow$'': If $M$ accepts,
$\vec{s}=I_{G'}(\vec{s})$ and hence $\vec{s}=\vec{v}'$. It remains
to prove that if $M$ accepts, $\vec{z}=\vec{v}$ holds. Suppose $M$
accepts but $\vec{z}\neq\vec{v}$. If $\vec{z}\neq\vec{v}$, then
$|z(i)-v(i)|\geq\delta$ for at least one $i\in V$. Since $M$ accepts,
$|z(i)-s(i)|=|z(i)-v(i)'|<\delta/2$ for all $i\in V$. It follows
that $|v(i)-v(i)'|\geq\delta/2$ for at least one $i\in V$ which
contradicts the construction of $G'$. {}``$\Leftarrow$'': Suppose
$\vec{z}=\vec{v}$ and $\vec{s}=\vec{v}'$. Then obviously $\vec{z}=I_{G}(\vec{z})$,
$\vec{s}=I_{G'}(\vec{s})$ and $|z(i)-s(i)|<\delta/2$ for all $i\in V$
by construction of $G'$. Hence M accepts. \end{proof}

\textbf{\emph{Algorithms:}} 

All algorithms \cite{BjorklundV05,Ludwig95,Condon93,Kumar04,Somla05,FilarSTV91,SzepesvL96}
suggested to date for the problem of finding a solution to \eqref{eq:optimalvaluevector}
have exponential time complexities. The most intuitive of those operates
on the basis of the iterative update rule $\vec{v}_{i+1}=I_{G}(\vec{v}_{i})$
and is called successive approximation or value iteration algorithm.
As already mentioned, this algorithm is guaranteed to converge to
the correct solution only if $G$ is stopping. In her paper about
algorithms for simple stochastic games, Condon \cite{Condon93} presents
a {}``worst case'' example for the successive approximation algorithm
in form of a special game graph, where the algorithm takes an exponential
number of updates until it finds the optimal value vector. 

So called strategy improvement algorithms try to iteratively improve
an initial pair of strategies until convergence. A particularly simple
algorithm of this class is the one of Hoffman \& Karp \cite{Kumar04},
for which a worst case running time of $O(2^{n}/n)$ is established.
Björklund \& Vorobyov's \cite{BjorklundV05} randomized algorithm
dating from 2005 has a worst case running time of $O(2^{^{\sqrt{n\cdot log(n)}}})$,
which as of the authors knowledge, is the best result obtained to
date.

\subsubsection*{\emph{SSG-VALUE{*}} \textmd{(decision)}}

\begin{longtable}{|ll|}
\hline 
Input:&
Simple stochastic game $G$, $\alpha\in\Omega_{n}$ \tabularnewline
Question:&
Is the value of $G>\alpha$?\tabularnewline
\hline
\end{longtable}

\textbf{\emph{Complexity}} \textbf{}\textbf{\emph{class:}} $NP\cap coNP$

\emph{SSG-VALUE{*}} is a straightforward extension of the \emph{SSG-VALUE}
Problem, which is defined by $\alpha=1/2$. Using the same terminology
as for the previous problem, a nondeterministic Turing-machine $M$
for \emph{SSG-VALUE{*}} first guesses one vector from the set $\Omega_{n'}^{n'}$
to be the optimal value vector of the $1/2^{9n}$-stopping game $G'$
of $G$, where $n'=9n|E|+n$. If we denote this vector by $\vec{s}$
and define that $M$ accepts $\vec{s}$ if $\vec{s}=I_{G'}(\vec{s})$
and $s(start)>\alpha$, then $M$ accepts $\vec{s}$ if and only if
$value>\alpha\wedge\vec{s}=\vec{v}'$. It should also be clear that
$M$ comes to a conclusion in time polynomial in the size of $G$.

\begin{proof}
 {}``$\Rightarrow$'': If $M$ accepts $\vec{s}$, then $\vec{s}=I_{G'}(\vec{s})$
and therefore $\vec{s}=\vec{v}'$. It remains to show that if $M$
accepts $\vec{s}$, $value>\alpha$. Suppose $M$ accepts $\vec{s}$
but $value\leq\alpha$. If $value\leq\alpha$ then $value'=s(start)\leq\alpha$
since the stopping game $G'$ always has lower value by construction.
Hence $M$ cannot accept $\vec{s}$. {}``$\Leftarrow$'': Suppose
$value>\alpha$ and $\vec{s}=\vec{v}'$. Then $value\geq\alpha+\delta$
and since, by construction of $G'$, $|value-s(start)|\leq\delta/2$,
it follows that $s(start)>\alpha$. Since also $\vec{s}=I_{G'}(\vec{s})$,
M accepts $\vec{s}$. Applying a small modification to $M$ by letting
it accept if $s(start)\leq\alpha$, $M$ can obviously also decide
the complement of \emph{SSG-VALUE{*}} . \end{proof}

\textbf{\emph{Algorithms}}: 

The author is not aware of any algorithm specially tailored for \emph{SSG-VALUE{*}},
instead, algorithms for the more general \emph{SSG-OVV} are used to
solve \emph{SSG-VALUE{*}}. Though theoretically, algorithms for \emph{SSG-VALUE{*}}
could exploit the circumstance that -- depending on the topology of
$G$ -- not all optimal vertex values of $G$ would need to be computed
in order to solve the problem, as we are mainly concerned with the
worst case behavior of such algorithms, the case where the value of
$G$ depends on its whole game graph must be assumed. Therefore, the
value vector of $G$ must be computed after all.

\section{Conclusion and Open Problems}

We have seen that the most interesting and also most difficult simple
stochastic game problem, that of computing the optimal value vector,
is hard to solve. However, restricting the input to simple stochastic
games that lack one vertex kind, we observed that the same problem
becomes tractable and can be solved rather efficiently. Another result
we obtained was that the value vector of reduced games can be computed
in polynomial time, i.e that given a simple stochastic game and a
pair of player strategies, the question about the winning-probabilities
of the players is efficient answerable. 

Sadly, we were not able to show polynomial time equivalence of \emph{SSG-OVV}
and \emph{SSG-VALUE{*}} but we would still like to know what the decision
equivalent of \emph{SSG-OVV} is. Another major open problem is to
show completeness of \emph{SSG-VALUE} (\emph{or} \emph{SSG-OVV}) for
a specific complexity class, thereby specifying the problem's exact
complexity. In a last word, Condon expressed the possibility that
finding an algorithm that seperates simple stochastic games with low
value $(\alpha<0,25)$ from those with high value $(\alpha>0,75)$
might be of significance in solving the master problem: \emph{SSG-VALUE}
\emph{$\in P$}?

\section*{\appendix Appendix A}

\begin{algorithm}[H]

\caption{optimalValueVector$(G)$}

\begin{algor}
\item [{{*}}] \textbf{input} simple stochastic game $G=(V,E)$
\item [{{*}}] \textbf{output} optimal value vector \textbf{}$\vec{v}$
of $G$
\item [{{*}}] \textbf{require} $V_{avg}=\emptyset$
\item [{{*}}] \textbf{begin}
\item [{{*}}] $D=\{ n-1,n\},$ $\vec{v}=\vec{0}$
\item [{repeat}]~
\item [{for}] $i\in V\setminus D$
\item [{if}] $i$ is a max vertex with a 1-valued child in $D$ 
\item [{{*}}] $D=D\cup\{ i\},$ $v(i)=1$
\item [{elseif}] i is a max vertex with two 0-valued children in $D$
\item [{{*}}] $D=D\cup\{ i\},$ $v(i)=0$
\item [{elseif}] i is a min vertex with a 0-valued child in $D$
\item [{{*}}] $D=D\cup\{ i\},$ $v(i)=0$
\item [{elseif}] i is a min vertex with two 1-valued children in $D$
\item [{{*}}] $D=D\cup\{ i\},$ $v(i)=1$
\item [{endif}]~
\item [{endfor}]~
\item [{until}] $D$ is static
\item [{{*}}] \textbf{return} $\vec{v}$
\item [{{*}}] \textbf{end}
\end{algor}

\end{algorithm}

\bibliographystyle{plain}
\bibliography{papers/biblio}

\end{document}